\documentclass[aps,twocolumn,showpacs,preprintnumbers,amsmath,amssymb]{revtex4}

\usepackage{graphicx}
\usepackage{dcolumn}
\usepackage{bm}
\usepackage{amsmath}

\begin{document}


\title{Spectral Tripartitioning of Networks}

\author{Thomas Richardson$^1$}
\author{Peter J. Mucha$^{1,2}$}
\homepage{http://netwiki.amath.unc.edu}
\email{mucha@unc.edu}
\author{Mason A. Porter$^3$}
\affiliation{%
$^1$Carolina Center for Interdisciplinary Applied Mathematics, Department of Mathematics,\\ University of North Carolina, Chapel Hill, NC 27599-3250, USA;}%
\affiliation{%
$^2$Institute for Advanced Materials, Nanoscience and Technology\\
University of North Carolina, Chapel Hill, NC 27599;}%
\affiliation{%
$^3$Oxford Centre for Industrial and Applied Mathematics, Mathematical Institute,\\ University of Oxford, OX1 3LB, United Kingdom}%

\date{\today}

\begin{abstract}

We formulate a spectral graph-partitioning algorithm that uses the two leading eigenvectors of the matrix corresponding to a selected quality function to split a network into three communities in a single step.  In so doing, we extend the recursive bipartitioning methods developed by Newman [Proc.\ Nat.\ Acad.\ Sci.\ {\bf 103}, 8577 (2006); Phys.\ Rev.\ E {\bf 74}, 036104 (2006)]
to allow one to consider the best available two-way and three-way divisions at each recursive step. We illustrate the method using simple ``bucket brigade'' examples and then apply the algorithm to examine the community structures of the coauthorship graph of network scientists and of U.~S. Congressional networks inferred from roll-call voting similarities.

\end{abstract}

\maketitle


\section{Introduction} \label{sec:intro}

Networks, or ``graphs,'' provide a powerful representation for the analysis of complex systems of interacting entities.  This framework has opened a large array of analytical and computational tools, and the study of networks has accordingly become pervasive in sociology, biology, information science, and many other disciplines \cite{str01,Newman2003,mendes,cald,newmanphystoday}.  The simplest type of network---an unweighted, undirected, unipartite graph---consists of a collection of nodes (representing the entities) that are connected by edges (representing the ties/links).  Important generalizations include weighted edges (ties with different strengths), directed edges (links from one node to another without reciprocation), and signed edges (e.g., ties interpreted as good or bad).  Many networks in applications are also bipartite, with two types of nodes and ties that always connect nodes of one type to those of the other \cite{Newman2003}.

To better understand the structural and functional organization of
networks, it is useful to develop computational techniques to
detect cohesive sets of nodes called ``communities," which can be identified
as groups of nodes that have stronger internal ties than they have to
external nodes
\cite{santo,santobig,satu07,structpnas,Newman2003,commreview,ourreview}.
The larger density of intra-community edges versus inter-community
edges, relative to what one might expect at random, has been shown
in many cases to correspond to increased similarity or association
among nodes in the same community.  For example, communities in
social networks might correspond to circles of friends, and
communities in the World Wide Web might correspond to pages on
closely-related topics.  Over the last seven years, the detection of communities has become a particularly active and important area of network science \cite{ourreview,santobig,santo,satu07}.
Community-detection efforts have yielded several
striking successes, offering insights into college
football ranking systems \cite{structpnas,bcs}, committee
\cite{congshort,conglong,conggallery} and cosponsorship \cite{yan}
collaborations in the United States Congress, functional motifs in
biological networks \cite{milo,amaral}, social structures in
cellular-phone conversation networks \cite{palla07}, social
organization in collegiate friendship networks \cite{facebook}, and
more.

Available algorithms to detect communities take a variety of forms, including linkage clustering \cite{cluster}, betweenness-based methods \cite{structpnas,walkbetween}, local techniques \cite{bagrow,vicsek,clausetlocal,santo2}, and spectral partitioning \cite{newman2006pnas,newman2006pre}. Some of these approaches can be cast as computational heuristics for the optimization of quality functions, such as the global quantity of ``modularity'' (and variants thereof) \cite{structmix,structeval,markfast,spinglass} or local quantities that more intimately measure the roles of links both between individual nodes and between/within individual communities \cite{santo,satu07}.  Most community-detection algorithms can be classified into one of three categories: recursive partitioning, local agglomeration, or direct calculation into a final number of communities.  The last approach tends to be computationally expensive, whereas the first two can misappropriate nodes and typically require some heuristic choices in the development of the algorithm.  In many community-detection methods, the constructed communities can also be layered in a hierarchical fashion, though the resulting hierarchy might depend strongly on the algorithm employed rather than just on the hierarchy of communities in the actual network.  Some methods also allow one to study overlapping communities \cite{clausetlocal,bagrow,vicsek,santo2,wiggins08}, though in the present discussion we consider only partitioning into nonoverlapping communities.

Given the expanse and constant, rapid advances in the area of community
detection, we do not endeavor to more fully catalog or compare the
numerous methods that are now available in the literature.  Such discussions are now available in several review articles \cite{santo,santobig,commreview,ourreview}.  Starting
from the recognition that using spectral partitioning to optimize
modularity and other similar quality functions is one of the (many)
available and preferred means for identifying communities, we focus
on that family of methods for the remainder of this paper.  We present a fundamental extension to this class of methods and demonstrate the resulting improvement with some examples.

In traditional spectral partitioning, which arose most prominently in the development of algorithms for parallel computation, one relates network properties to the spectrum of the graph Laplacian matrix \cite{fiedler,pothen}.  In the simplest such procedure, one starts by partitioning a network into two subnetworks of specified size.  One then examines the resulting subnetworks and further divides them if desired, continuing this procedure recursively for as many divisive steps as desired.  After community detection became prominent in network science, spectral methods were generalized to include some algorithms for community detection, including work that includes steps that go beyond two-way splits \cite{capocci,donetti}.  In a recent pair of papers \cite{newman2006pnas, newman2006pre}, Newman reformulated the idea of maximizing modularity as a spectral partitioning problem by constructing a \emph{modularity matrix} and using its leading eigenvector to spectrally partition networks into two subnetworks.  He then applied recursive subdivision until no further divisions improved modularity, resulting in a final collection of communities whose number and sizes need not be specified in advance.  This is an important feature for the study of communities in just about every social, biological, and information network application, as there is often no way to know the numbers and sizes of communities in advance.

Given the NP-complete nature of modularity optimization \cite{np}, the polynomial-time spectral algorithm does not guarantee a global optimum. Indeed, Ref.~\cite{newman2006pre} includes a simple example of eight vertices connected together in a line, in which the best partition found by recursive bipartition consists of 2 groups of 4 nodes each, whereas the exhaustively-enumerated optimum partition consists of 3 communities.  The latter partition cannot be obtained by this recursive bipartition method because the initial split occurs in the middle of the line.  Reference~\cite{newman2006pre} also explores some possibilities for using multiple leading eigenvectors of the modularity matrix but does not pursue the idea in detail beyond a two-eigenvector method for bipartitioning.

In the present paper, we provide a valuable extension of
the spectral partitioning methods for community detection in which we use two
eigenvectors to tripartition a network (or subnetwork) into three
groups in each step. This method can be combined with the one- and
two-eigenvector bipartitioning methods to more thoroughly explore
promising partitions for computational optimization of the selected
quality function (either modularity or other choices) with only limited additional computational cost.
In developing this tripartitioning extension, we employ a modified
Kernighan-Lin (KL) algorithm \cite{kl} (see also Refs.\ \cite{newman2006pnas,newman2006pre}; other modifications of KL are also possible \cite{blondel,kreso}).  We illustrate the
resulting spectral method for community detection with the same
nodes-in-a-line ``bucket brigade'' networks that are not always
optimally partitioned by recursive bipartitioning. As examples, we then
apply the method to similarity networks constructed from U.~S.
Congressional roll call votes \cite{pr,voteview,waugh} and to the
graph of network scientist coauthorships \cite{newman2006pre}. In so
doing, we include the important consideration that quality functions
other than the usual modularity measure can be similarly used in such spectral partitioning (without otherwise altering the algorithm in any way) provided that they can be cast in a similar matrix form.

The rest of this paper is organized as follows.  In Section \ref{sec:spectral}, we review the existing formulation for spectral community detection.  In particular, we discuss how to recursively bipartition a network using either the leading eigenvector or the leading pair of eigenvectors of the modularity matrix.  In Section \ref{sec:theory}, we present a theory (and polynomial-time algorithm) that extends these ideas to three-way subdivision (\emph{tripartitioning}) using the leading eigenvector pair.  In Section \ref{sec:implement}, we provide a faster implementation of this procedure in which we employ a restricted consideration of the possible cases and then leverage KL iterations to identify high-quality partitions.  We subsequently present several examples in Section \ref{sec:examples}, highlighting situations in which allowing one to choose either two-way or three-way splits at each step in the recursion procedure results in higher-modularity partitions than recursive bipartitioning alone.  Finally, we summarize our results in Section \ref{sec:conclusions}.

\section{Review of Spectral Partitioning Using Modularity}
\label{sec:spectral}

In this study, we largely focus on the quality function known as modularity \cite{structmix,structeval,markfast}, which we attempt to maximize for a given undirected network via spectral partitioning.  We stress that our methods can be used with \textit{any} quality function that can be cast in a similar matrix form.  (We illustrate one such example in Section \ref{sec:examples}.)  Because the novel algorithms we present both extend and interface with existing spectral partitioning methods, it is necessary to review the essential elements of Refs.~\cite{newman2006pnas,newman2006pre} that recur in the subsequent presentation of our tripartitioning scheme.

Starting from the definition of a network in terms of its nodes and (possibly weighted) edges, we denote the strengths of connections using a symmetric adjacency matrix $\mathbf{A}$, whose components $A_{ij} = A_{ji}$ codify the presence and strength of connection between nodes $i$ and $j$. In an unweighted network, each $A_{ij}$ component has a binary $\{0,1\}$ value. Given some partition of the network, the modularity $Q$ can be used to compare the total weight of intra-community connections relative to the weight that one would expect on average in some specified null model \cite{newman2006pre}:
\begin{equation}
    	Q = \frac{1}{2m}\sum_{g = 1}^c \sum_{i,j\in G_g}B_{ij} = \frac{1}{2m}\mathrm{Tr}(\mathbf{S}^T\mathbf{BS})\,, \label{modularity}
\end{equation}
where $\mathbf{B}$ is called the modularity matrix, $m = \frac{1}{2}\sum_{ij}A_{ij}$ is the total edge weight in the network, $B_{ij} = A_{ij}-P_{ij}$ for a selected null model matrix $\mathbf{P}$ with elements $P_{ij}$, the sum over $g$ runs over the $c$ communities in the specified partition, and the set of vertices $G_g$ comprises the $g$th community. The $n$-by-$c$ ``community-assignment matrix'' $\mathbf{S}$ encodes the non-overlapping assignment (``hard partitioning" \cite{santobig}) of each of the $n$ nodes to the $c$ communities:
\begin{equation}
    S_{ig} = \left\{
        \begin{array}{c l}
              1\,, & \mbox{if node $i$ belongs to community $g$,}\\
              0\,, & \mbox{otherwise.}\\
        \end{array}\right.
    \label{Smatrix}
\end{equation}

The most commonly studied null model for unipartite networks recovers the Newman-Girvan definition of modularity \cite{structeval}, which is obtained by considering the ensemble of random graphs with independent edges conditional on having the same expected strength distribution as the original network.  This gives $P_{ij} = k_ik_j/(2m)$, where $k_i=\sum_j A_{ij}$ is the total edge weight (``strength'') of the $i$th node (equivalent to its degree in the unweighted case).  One then constructs the modularity matrix $\mathbf{B}$ by subtracting these expected connection strengths from the connection weights in $\mathbf{A}$.

Importantly, modularity is not the only quality function that can be
cast in the form of \eqref{modularity}. In cases that are seemingly
closer to uniform random graphs, such as those encountered in the
study of network tie strengths inferred from similarities (as
occurs, for example, when studying voting patterns in roll calls
\cite{waugh}), a uniform $P_{ij} = p$ null model might be an appropriate
alternative.  Moreover, modularity does not always provide a suitable resolution of a network's community structure. One means of overcoming this
deficiency, which gives the ability to examine a network at different resolution levels, is to multiply the null model by a resolution parameter, yielding $P_{ij} = \gamma k_ik_j/(2m)$ \cite{spinglass,resolution}.  (This is related to the sizes of communities obtained using random walk processes over different
time intervals \cite{barahona}.)  Another means of introducing a
resolution parameter is by addition of (possibly signed) self-loops
along the diagonal of a modified adjacency matrix and carrying the resulting changes into the usual modularity null model \cite{arenas08}.  One can also consider directed networks with appropriate null models by using the symmetric part of $\mathbf{B}$ \cite{leicht08}.

In the present work, we do not restrict ourselves by making any assumptions about the null model beyond its representation in terms of a $\mathbf{B}$ matrix in \eqref{modularity}. However, the application to large, sparse networks requires that this matrix have sufficient structure to enable efficient computation of the product $\mathbf{B}\mathbf{v}$ for arbitrary vectors $\mathbf{v}$. All of the quality functions mentioned above have this property. Indeed, we will include consideration of the additional self-loops model in the Congressional roll call example.

We now review the bipartitioning of nodes into two communities (not necessarily of equal size) using the leading eigenvector of $\mathbf{B}$.  In calculating $Q$ (or other relevant quality function), one can replace the role of the $n$-by-$2$ community assignment matrix $\mathbf{S}$ by a single community vector $\mathbf{s}$ with components $s_i = \pm 1$ indicating the assignment of node $i$ to one of two groups.  Because $\frac{1}{2}(s_i s_j + 1)$ eqiuvalently indicates whether nodes $i$ and $j$ have been placed in the same community, it follows that one can attempt to optimize $Q_s = \frac{1}{4m}\mathbf{s}^T\mathbf{Bs}$. If the null model maintains $\sum_{ij} B_{ij} = 0$ identically, then $Q_s = Q$.  For other null models, the difference between $Q$ and $Q_s$ is a constant specified by the adjacency matrix and null model, so optimization of $Q_s$ is equivalent to optimization of $Q$. Consequently, the leading eigenvector $\mathbf{u}_1$ (associated with the largest positive eigenvalue) of $\mathbf{B}$ gives the maximum possible value of $\mathbf{v}^T\mathbf{Bv}$ for real-valued $\mathbf{v}$.  Heuristically, one thus expects the community assignment indicated by $s_i = \mathrm{sgn}([\mathbf{u}_1]_i)$ to give a large value of modularity.  (When $[\mathbf{u}_1]_i=0$, one can set $s_i=\pm 1$ according to whichever choice gives larger modularity.  If KL iterations are going to be used later to improve community assignments, it is sufficient to use a simpler rule of thumb.)

Similar ideas can be used to bipartition using multiple leading eigenvectors \cite{newman2006pre}. Without specifying all of the details here, the $p$-eigenvector approach starts by the selection of $n$ node vectors $\mathbf{r}_i$ whose $j$th components ($j \in \{1,\dots, p\}$) are determined by
\[
    [\mathbf{r}_i]_j = \sqrt{\beta_j-\alpha}\, U_{ij}\,,
\]
where $\beta_j$ is the eigenvalue of $\mathbf{u}_j$ (i.e., the $j$th ordered eigenvector of $\mathbf{B}$), $\mathbf{U} = (\mathbf{u}_1\vert \mathbf{u}_2\vert \cdots)$, and the constant $\alpha \leq \beta_p$
is related to the approximation for $Q$ obtained by using only the first $p$ vectors (proceeding using as many positive $\beta_j$ as desired).  For computational convenience, we hereafter set $\alpha=\beta_n$. The modularity is then approximated by the relation
\begin{equation}
    Q \approx \widetilde{Q} \equiv n\alpha + \sum_{g=1}^c\left\vert\mathbf{R}_g\right\vert^2\,,
\label{eq:QR}
\end{equation}
where the sum is over the number of communities ($c$) and the contribution of each community is given by the magnitude of the associated community vector $\mathbf{R}_g = \sum_{i\in G_g} \mathbf{r}_i$.  Important to further developments in Ref.~\cite{newman2006pre} and by us below is that the assignments of nodes to communities that maximize the sum in (\ref{eq:QR})
require that $\mathbf{R}_g\cdot\mathbf{r}_i>0$ if node $i$ has been assigned to community $g$. By contrast, if $\mathbf{R}_g\cdot\mathbf{r}_i\leq 0$, then simply reassigning node $i$ from community $g$ to its own individual group increases the modularity approximation \eqref{eq:QR} by
\[
\Delta\widetilde{Q} =
\vert\mathbf{R}_g-\mathbf{r}_i\vert^2 + \vert\mathbf{r}_i\vert^2 - \vert\mathbf{R}_g\vert^2 = 2\vert\mathbf{r}_i\vert^2 - 2\mathbf{R}_g\cdot\mathbf{r}_i > 0\,.
\]
Similarly, all pairs of communities $\{g,h\}$ in the partition that optimizes \eqref{eq:QR} must be at least $90^\circ$ apart (i.e., $\mathbf{R}_g\cdot\mathbf{R}_h<0$), because the change in $\widetilde{Q}$ from merging two communities is
\[
    \vert\mathbf{R}_g+\mathbf{R}_h\vert^2 -\left(\vert\mathbf{R}_g\vert^2+\vert\mathbf{R}_h\vert^2\right) = 2\mathbf{R}_g\cdot\mathbf{R}_h\,.
\]
Because the maximum number of directions more than $90^\circ$ apart that can exist simultaneously in a $p$-dimensional space is $p+1$, the $p$-dimensional representation of the vertices and communities restricts the spectral optimization of \eqref{eq:QR} to a partitioning (in a single step) into at most $p+1$ groups.

\begin{figure}
\centering
\includegraphics[angle=0, width=.5\textwidth]{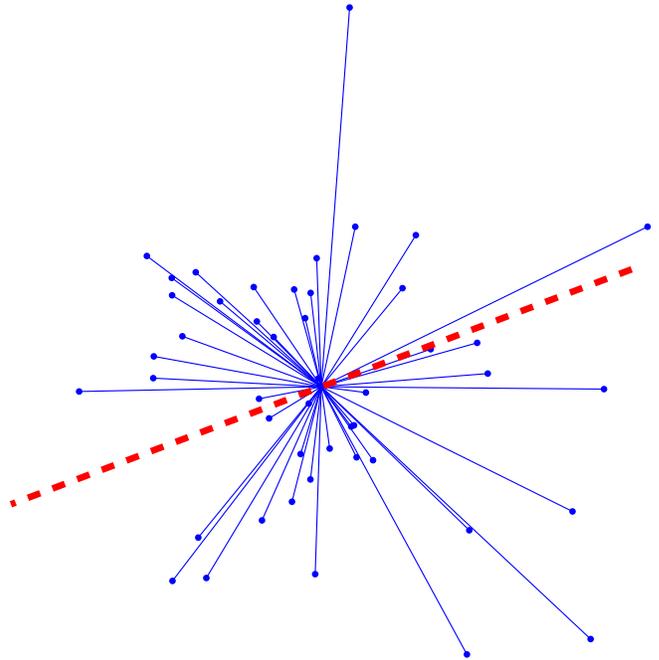}
\caption{(Color online) Bisection in two dimensions (i.e., using the leading pair of eigenvectors of $\mathbf{B}$) for a small network. Solid (blue) lines with dots represent node vectors. The dotted (red) line represents a selected cut line in the plane.  All nodes on one side of the line are assigned to one community and all nodes on the other side are assigned to the other community. Rotating this line about the origin yields the set of possible planar bipartitions.  (This figure is adopted from one in Ref.~\cite{newman2006pre}.)}
\label{fig:cutplane}
\end{figure}

Leveraging the above geometric constraints, the bipartition-optimizing \eqref{eq:QR} must be equivalent to some bisection of the node vector space by a codimension-one hyperplane that separates the vertices of the two communities.  Computational use of this observation requires efficient enumeration of the allowed partitions.  For instance, in the two-eigenvector planar case ($p = 2$, assuming at least two positive eigenvalues), only $n/2$ distinct partitions are allowed by the geometric constraints.  As shown in Fig.~\ref{fig:cutplane}, each permissible partition is specified by bisecting the plane according to a cut line that passes through the origin \cite{newman2006pre}. The algorithm for spectral bipartitioning by two eigenvectors then proceeds by considering each of the $n/2$ allowed partitions and selecting the best available one.

Recursive bipartitioning can then be used to split a network into as many communities as desired or until it can no longer further improve the value of the quality function. This recursive subdivision must be done with a generalized modularity matrix in order to properly account for the contribution to modularity
from further subdivisions of subnetworks \cite{newman2006pre}. Specifically, the change in modularity given by subdivision of the $n_G$-node group $G$ into $c_G$ smaller groups specified by $S_{ig}$ can be recast as a similar spectral partitioning problem with the $n_G\times n_G$
generalized modularity matrix $\mathbf{B}^{(G)}$ taking the place of $\mathbf{B}$.  Its elements, indexed by the node labels $i \,,j \in G$, are specified in terms of $\mathbf{B}$ by
\begin{equation}
	B^{(G)}_{ij} = B_{ij} - \delta_{ij}\sum_{l\in G} B_{il}\,. \label{eq:genB}
\end{equation}
One can then implement such subdivisions recursively until modularity can no longer be increased with any additional partitioning.

As an aside, we remark that spectral bipartitioning using any null model for which the rows of $B_{ij}$ (or its symmetric part) sum to zero has the additional property
that $\mathbf{R}_1 = -\mathbf{R}_2$ because the zero row sums guarantee by eigenvector orthogonality that $\sum_k\mathbf{R}_k = \sum_i\mathbf{r}_i = 0$. Meanwhile, the
generalized modularity matrix $\mathbf{B}^{(G)}$ for recursive subdivision has zero row sums by
construction. The absence of this property for the initial division for more general
null models does not affect the implementation of either any of the algorithms
described above or of our recursive tripartitioning algorithm that we present below.

Finally, we reiterate that individual partitioning steps and the
recursive implementation of such steps can be employed as a
computational heuristic for optimizing \textit{any} quality function
(not just modularity) that can be written in a matrix form similar
to \eqref{modularity}.  Naturally, this approach is not the only available heuristic for optimizing a specified quality function \eqref{modularity} or its vector approximation \eqref{eq:QR} (see, e.g., the many references mentioned earlier). For instance, Ref.~\cite{vectorwang} uses \eqref{eq:QR} as the basis of
an eigenvector-ordered vector-partitioning algorithm that uses
bisection in each coordinate as the starting point for collecting
nodes into the geometrically-constrained number of groups (which, we recall, is one more
than the dimension of the ambient space).  In this sense, their algorithm has some similarities in two dimensions to the one we describe below. 
 However, we take a different approach: We first
establish in Section \ref{sec:theory} the relevant inequalities and
geometry of the problem of tripartitioning (which requires generalizing the
constraints we reviewed above) and then propose a divide-and-conquer
implementation strategy in Section \ref{sec:implement}.  We
subsequently apply our techniques to an illustrative example and
real-world networks in Section \ref{sec:examples}.

\section{Eigenvector-Pair Tripartitioning: Theory} \label{sec:theory}

\begin{figure}
\centering
\includegraphics[angle=0, width=.5\textwidth]{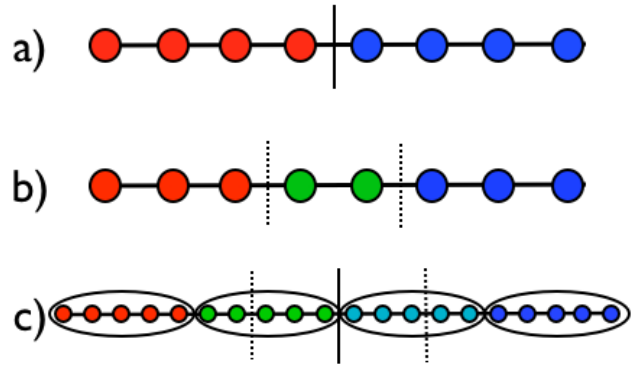}
\caption{(Color online) The 8-node bucket brigade network discussed in Ref.~\cite{newman2006pre} and a similar 20-node bucket brigade network. Solid lines between nodes represent connections. Panel (a) shows the best initial bipartition (solid vertical line) of the 8-node network (no further subdivision increases modularity). Panel (b) shows the modularity-maximizing partition (dotted lines) of the same network.  Note that the three-community partition in (b) cannot be obtained directly from (a) via subsequent partitioning.  Panel (c) shows the maximum modularity partition of the 20-node network into four communities (indicated by ovals around nodes and also by colors online), compared with the partition obtained via the best initial three-way division (vertical dotted lines), which is larger than that obtained by the initial bipartition (vertical solid line).  In this case, the four-way partition cannot be obtained from recursive partitioning of the initial three-way division, but it is obtained from recursive partitioning of the bipartition.}
\label{fig:lines}
\end{figure}

Recursive network bipartitioning, combined with subsequent KL iterations
(which we describe and use in Section \ref{sec:implement}), rapidly
produces high-quality partitions.  However, given the algorithms' polynomial
running times, such values are not guaranteed to be optimal. Indeed,
Ref.~\cite{newman2006pre} describes a simple case of an 8-node line
segment (bucket brigade) network in which these algorithms miss the
optimal-modularity partition.  The recursive bipartitioning procedure initially bisects the network into
two groups (see Fig.~\ref{fig:lines}a) and then terminates because no
further subdivision improves modularity.  In
contrast, the modularity-maximizing partition of this small test
network (which can be obtained by exhustive enumeration) consists of three communities (see
Fig.~\ref{fig:lines}b) that cannot be obtained by
subsequent subdivision of the initial bisecting split.  Figure~\ref{fig:lines} also shows a similar 20-node bucket brigade network 
that we will discuss in more detail in Section \ref{sec:implement}.

\begin{figure}
\centering
\includegraphics[angle=0, width=.5\textwidth]{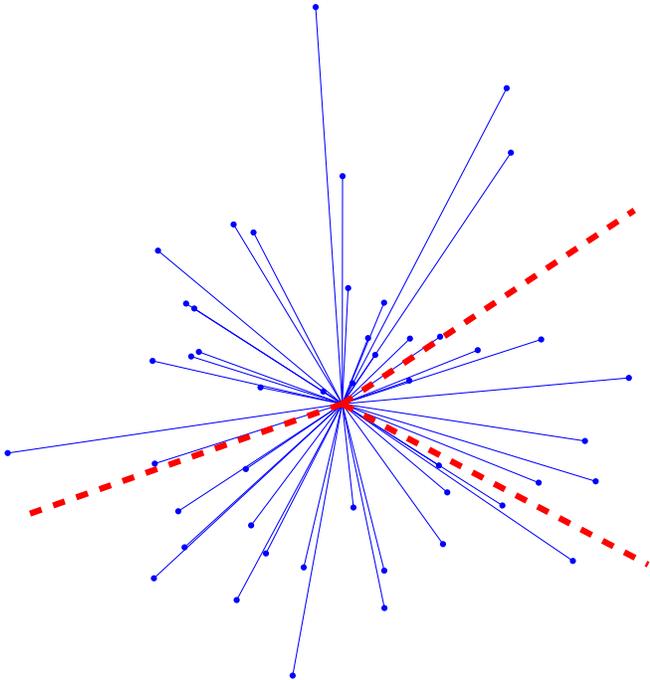}
\caption{(Color online) Tripartitioning with vertex vectors from the leading pair of eigenvectors of $\mathbf{B}$ for a small network.  Solid (blue) lines with dots represent node vectors, which have been rescaled according to the observed standard deviation of each component. The dotted (red) rays border a selected set of wedges in the plane, where each wedge indicates the set of nodes assigned to one group.  One obtains the tripartitions allowed by the planar geometric constraints by rotating the boundaries of these wedges about the origin.}
\label{fig:wedges}
\end{figure}

Motivated by the above example, the efficient two-vector bipartitioning
algorithm, and the geometric constraints that limit a $p$-dimensional node vector space representation to at most $p+1$ communities (as described
in Section~\ref{sec:spectral}), we consider whether a
similarly-efficient mechanism exists for dividing the plane of node
vectors into three groups in a single partitioning step. We start
by considering the generalization of the cut line (illustrated in Fig.~\ref{fig:cutplane}) to a set of three non-overlapping wedges that 
fill the plane (as shown, e.g., in Fig.~\ref{fig:wedges}). That is,
instead of a single cut line that intersects the origin and bisects
the plane, we ask whether the planar tripartitions that are
geometrically permitted by \eqref{eq:QR} are equivalent to finding three rays emanating from the
origin, with each (non-overlapping) wedge between these rays
specifying the vertices of a group. That is,
vertices are assigned to a community if they lie between the rays
denoting that community's boundaries.

As we now show, such non-overlapping wedges \footnote{Note that our
use of the term ``overlap'' in this specific setting is distinct
from the common terminology of ``overlapping communities'' that allows
nodes to belong to multiple groups \cite{santobig,ourreview}.} describe
the geometric constraint that vertices whose vectors are located
inside one wedge cannot be assigned to a community associated with
another wedge in the partition that maximizes \eqref{eq:QR} in the
plane. Consider two planar community vectors, $\mathbf{R}_1$ and
$\mathbf{R}_2$ (with $\mathbf{R}_1\cdot\mathbf{R}_2 < 0$, as
discussed in Section \ref{sec:spectral}).  Also suppose that there
is a node vector $\mathbf{r}_0$ within $90^\circ$ of each community
vector, so that $\mathbf{r}_0\cdot\mathbf{R}_g > 0$ for
$g\in\{1,2\}$. Without loss of generality, we introduce a second
node vector $\mathbf{r}_1$ in the (smaller-angle) region between
$\mathbf{r}_0$ and $\mathbf{R}_2$. The proof of non-overlapping
wedges for the optimization of \eqref{eq:QR} then reduces to showing
that if $\mathbf{r}_1$ is assigned to the $\mathbf{R}_1$ group, then
$\mathbf{r}_0$ must be assigned to $\mathbf{R}_1$ as well. That is,
in the event that $\mathbf{r}_0$ has been assigned to
$\mathbf{R}_2$, the $\Delta\widetilde{Q}_{01}$ improvement in moving
node $0$ to group $1$ should be positive. Similarly, the condition
that $\mathbf{r}_1$ is (correctly) assigned to the $\mathbf{R}_1$
group requires both that $\mathbf{r}_1\cdot\mathbf{R}_1>0$ and that
$\Delta\widetilde{Q}_{12}\leq 0$ for moving node $1$ to group $2$,
where
\begin{align*}
    \Delta\widetilde{Q}_{12} &= \vert\mathbf{R}_1-\mathbf{r}_1\vert^2 + \vert\mathbf{R}_2+  \mathbf{r}_1\vert^2 -\vert\mathbf{R}_1\vert^2 -\vert\mathbf{R}_2\vert^2\\
    &= 2\vert\mathbf{r}_1\vert\left\{\vert\mathbf{r}_1\vert - 2\vert\mathbf{R}_1\vert\cos{\theta_{11}} +    2\vert\mathbf{R}_2\vert\cos{\theta_{12}}  \right\}
\end{align*}
and $\cos{\theta_{vg}} = \mathbf{r}_v\cdot\mathbf{R}_g/(|\mathbf{r}_v||\mathbf{R}_g|)$. From the geometric ordering of the vectors specified above, it follows that $0 < \cos{\theta_{11}} < \cos{\theta_{01}} < 1$ and $0 < \cos{\theta_{02}} < \cos{\theta_{12}} < 1$.
The improvement in moving node $0$ to group $1$---that is, moving to a partition assignment by non-overlapping wedges---then becomes
\begin{align*}
    \Delta\widetilde{Q}_{01} &= \vert\mathbf{R}_1+\mathbf{r}_0\vert^2 + \vert\mathbf{R}_2-  \mathbf{r}_0\vert^2 -\vert\mathbf{R}_1\vert^2 -\vert\mathbf{R}_2\vert^2\\
    &= 2\vert\mathbf{r}_0\vert\left\{\vert\mathbf{r}_0\vert + 2\vert\mathbf{R}_1\vert\cos{\theta_{01}} -    2\vert\mathbf{R}_2\vert\cos{\theta_{02}}  \right\}\\
    &> 2\vert\mathbf{r}_0\vert\left\{\vert\mathbf{r}_0\vert + 2\vert\mathbf{R}_1\vert\cos{\theta_{11}} -    2\vert\mathbf{R}_2\vert\cos{\theta_{12}}  \right\}\\
    &= 2\vert\mathbf{r}_0\vert\left\{\vert\mathbf{r}_0\vert + \vert\mathbf{r}_1\vert - \frac{1}{2\vert  \mathbf{r}_1\vert}\Delta\widetilde{Q}_{12} \right\} > 0\,.
\end{align*}
Therefore, the optimal $\widetilde{Q}$ value in the plane must result from the assignment of nodes to groups equivalent to the specification of non-overlapping wedges.

Recall from Section \ref{sec:spectral} that the bisection of the plane into two halves, indicated by selecting a cut line, yields only $n/2$ distinct cases from which to select the best bipartition.  However, the corresponding enumeration for three-way division, in which one selects three rays that border wedges, leaves $O(n^3)$ distinct partitions in the plane that must be enumerated and evaluated.  While this provides an improvement over the original non-polynomial complexity of the partitioning problem, one still needs a more efficient heuristic for large networks to effectively employ such tripartitions (and subsequent subdivisions) at computational cost comparable with spectral bipartitioning.  We present such a heuristic in Section~\ref{sec:implement}.

\section{Eigenvector-Pair Tripartitioning: Fast Implementation}\label{sec:implement}

We accelerate the process of planar tripartitioning using a divide-and-conquer approach that reduces the number of considered configurations and, hence, the computational cost.
This approach yields a method that is computationally competitive with
two-eigenvector bipartitioning even for large networks.  In our
implementation, we start at a coarse level of considering the four
available tripartitions that can be obtained by unions of the
quadrants in the plane.  Before dividing these regions further, we rescale the individual coordinates of the plane according to the standard deviations of the observed components along each coordinate, keeping their original values for use in equation (\ref{eq:QR}).  This ad hoc rescaling makes the spatial distribution of vertex vectors more uniform and improves the efficiency of not only this coarse assignment but also subsequent refinements.  We then refine the quadrants by bisecting them into $w = 8$ wedge regions and individually consider each of the permissible unions of these new regions. If the best tripartition from $w = 8$ unions is better than that from quadrant unions, we bisect further to obtain $w = 16$ regions (which we then test), repeating this process of bisecting the pieces and finding their best union until the partition quality no longer improves (or no longer improves by some specified threshhold).

Obviously, this divide-and-conquer approach does not consider all $O(n^3)$ possible planar tripartitions. Indeed, at the stage in which the plane has been subdivided into $w = 2^j$ parts,
one considers at most $\binom{w}{3} = \frac{1}{6}w(w-1)(w-2)$ neighboring unions of the $w$ regions (including those already enumerated at smaller $w$ values). Some sets of nodes might repeat in this construction, and many such unions do not meet the full geometric constraints on group vectors, but it is easiest to code a search over all $\binom{w}{3}$ neighboring unions of the $w$ regions. The quality of the best partition obtained using such a small subset is of course unlikely to match the optimum obtained over all $O(n^3)$ allowed planar tripartitions.  However, as we demonstrate below and in Section \ref{sec:examples}, the resulting method does not appear to suffer from lower-quality
communities when combined with subsequent KL iterations (which we now describe).

The Kernighan-Lin (KL) iterative improvement scheme we use is a natural generalization \cite{newman2006pnas} of the original method \cite{kl} to the case where the sizes of communities is not specified in advance. A single KL iteration step consists of moving vertices one at a time from their assigned community to a different community such that each move provides the largest available increase (or smallest decrease) in the quality function,
subject to the constraint that each node is moved only once. That is,
as soon as a node is moved, it is removed from the list of those
available for consideration in upcoming moves in that step. These moves are selected independently of the geometric constraints placed on groups in the reduced eigenvector space; instead, the quality function is used directly to assess the value of each move.  After all $n$ nodes
have been moved precisely once each, one selects the best available partition from the $n$ that have been explored.  If that partition is of higher quality than the initial one, a new KL iteration step is started from this new best state.  Otherwise, the algorithm returns the initial partition as its final result.

Because KL iterations typically improve the best partitions constructed from the recursive spectral bipartitioning and tripartitioning algorithms, we recommend using them whenever possible.  The relatively high quality of the spectral partitions typically
manifest in a rapid convergence of the KL process. Moreover, the use of KL iterations is particularly important in light our divide-and-conquer approach for
reducing the number of planar tripartitions we consider. Specifically, we do not consider all of the $O(n^3)$ permissible planar tripartitions.  However, the
post-KL results do not appear to suffer in quality despite the
substantially-reduced number of configurations considered by our
divide-and-conquer strategy.  Consequently, the consideration of the full set of allowed planar wedges appears to be essentially unnecessary because the improvement obtained from such exceptional additional effort is overshadowed by the gains of the subsequent KL iterations.

As an example of the efficiency of this divide-and-conquer plus KL
approach, we consider the initial tripartition by modularity of the largest
connected component of the coauthorship graph of network scientists, which has $n = 379$ nodes and $914$ weighted edges \cite{newman2006pre}. (We will discuss this example in further detail in Section \ref{sec:examples}.)  The method does not identify a higher modularity among the allowed unions of $w = 64$ regions than that obtained at $w = 32$ (with $Q \doteq 0.5928$).  As a comparison, the best of the $O(n^3)$ allowed planar tripartitions has
modularity $Q \doteq 0.6175$, but comes at the cost of a greater than 200-fold increase in the number of configurations that must be considered.  In contrast, modularity is increased much more by using KL iterations after these two different tripartitions, with $Q \doteq 0.6354$ starting from the $Q \doteq 0.5928$ partition, and $Q \doteq 0.6349$ starting from the $Q \doteq 0.6175$ partition.  Observe that the higher post-KL modularity arises from the lower-modularity initial state, possibly due to an increased flexibility to move nodes between different groups; applying KL iterations starting from the best spectrally-identified tripartition here stays stuck near a local maximum.  Moreover, the best union at $w = 32$ in this example is only marginally ($1.4\%$) better than that obtained at $w = 16$.  One thus might make the algorithmic choice to only refine if the improvement in the quality function is better than some minimal threshold, because each refinement step of doubling $w$ increases the computational cost of this divide-and-conquer approach by a factor of about 8.  In the present work, we typically require the best modularity obtained with such a refinement to be at least $5\%$ better than previous values in order to justify the further doubling of $w$. We also stop the $w$-refinement process when $w$ becomes greater than the number of nodes in the network.

Based on the above discussion, we hypothesize that the reduced number of partitions considered have broadly sampled the gross global configuration possibilities sufficiently well so that the post-KL result should still be close in quality (and, as we have seen in our examples, even better in some cases) to that obtained after KL starting from the best of the permissible planar partitions.  We are intrigued by such thoughts but concerned that the complexity and detail of the local extrema of the selected quality function might not allow any general rigorous analysis.

We thus proceed with some representative examples using the aforementioned collection of spectral partitioning algorithms with subsequent KL iterations.  We obtained the results in Section \ref{sec:examples} using a recursive partitioning code that at each step selects the best partition from the available bipartitions and tripartitions
(using the fast implementations described above), followed by KL iterations starting from the point at which no further improvement in the quality function can be identified by spectral subdivision. We also generalize this partitioning algorithm further (as described in Ref.~\cite{yan}) by performing additional spectral subdivisions as if each group to be divided were the full network itself, isolated from the other groups in the partition (i.e., by constructing each $\mathbf{B}$ matrix directly from the adjacency submatrix restricted to one subnetwork at a time). Such steps decrease the global quality of the resulting partitions, even though they increase the quality of partitioning each individual subnetwork in isolation.  Despite the extra computational cost of this \textit{subnetwork-restricted partitioning} procedure, we have found that using KL iterations after such extended partitioning sometimes results in a higher global quality in the final partition. Unsurprisingly, these KL iterations correctly tend to merge a large number of the groups obtained from subnetwork-restricted partitioning during its search for the highest quality partition.

Obviously, one can devise many different variants on the above ideas, and the efficacy of one's results may vary (with better performance for some choices in specific examples). Additionally, one can significantly accelerate the spectral partitioning steps if the optimization in each step is based on the summed magnitude of the group vectors, approximating quality in \eqref{eq:QR}, by efficiently updating group vectors from one considered configuration to the next (as opposed to full recalculations).  When considering the full modularity (or other quality function) for different partitions, one can similarly accelerate the spectral steps using direct calculations of the differences in modularity between the configurations under consideration.

\section{Examples} \label{sec:examples}

Maintaining our focus on the utility of considering tripartitioning steps in recursive spectral partitioning with $\mathbf{B}$ eigenvectors, we proceed to consider example networks using the two implementations described in Section \ref{sec:implement} (with and without subnetwork-restricted partitioning).  We specifically include examples in which one can improve community-detection results by allowing tripartitioning steps and subnetwork-restricted modularity maximization. We draw our examples from different areas: the test ``bucket brigade'' networks of nodes connected to their nearest neighbors along a line segment, political networks constructed from voting similarities in U.~S. Congressional roll calls \cite{voteview}, and the coauthorship collaboration graph of network scientists \cite{newman2006pre}.

\subsection{Bucket Brigade Networks} \label{segments}

Because the 8-node line segment spurred our interest in spectral partitioning, we briefly use this network and its generalizations to further illustrate our results. The 8-node bucket brigade is the smallest nearest-neighbor line segment whose modularity-maximizing partition contains more than two communities.  (The 7-node bucket brigade has two-community and three-community partitions with equal modularity.)  However, the optimal partition (shown in Fig.~\ref{fig:lines}b) cannot be obtained from recursive bipartitioning, which terminates after the initial bisection of the network (see Fig.~\ref{fig:lines}a). In contrast, the tripartitioning method identifies the optimal partition in a single step. Subnetwork-restricted partitioning gives another means to identify the optimal partition through spectral partitioning with KL iterations. Specifically, applying this procedure to the initial bisection in Fig.~\ref{fig:lines}a further bisects each group, giving four groups of two nodes each, from which KL iterations merge two of these four groups on its way to the optimal three-group partition.

Yet another mechanism available to reach the optimal state is to allow the individual KL moves to place a node in a newly-created group of its own. If such moves are selected as tie-breakers over other moves that yield equal changes in modularity, then the optimal configuration of the 8-node bucket brigade can be obtained even in the absence of tripartitioning and subnetwork-restricted modularity maximization.
However, we warn that allowing the formation of such new groups as possible KL moves can drastically increase the number of groups under consideration (in some cases significantly beyond that obtained using subnetwork-restricted partitioning), so the increase in computational cost in generalizing KL iterations in this manner might not be worthwhile in all situations.
KL iterations can, of course, be generalized in other ways, such as by modifying the stopping condition or tie-breaking methods. In our results presented, we break ties randomly and do not allow new groups to be created in KL moves (except when explicitly indicated).

Contrasting the above, the 20-node bucket brigade, with a maximum modularity partition ($Q \doteq 0.5914$) of four communities with five nodes each (see Fig.~\ref{fig:lines}c), provides a cautionary illustration. Recursive bipartitioning by itself initially bisects the bucket brigade into two groups of 10 ($Q \doteq 0.4474$) and further subdivides each of those groups to correctly identify the optimal partition.  However, an initial tripartition (shown in Fig.~\ref{fig:lines}c), has higher modularity ($Q \doteq 0.5609$) than any initial bipartition, and recursive spectral steps that take the better result from bipartitioning and tripartioning terminate with a $Q \doteq 0.5720$ partition of sizes $n_g=\{4,3,6,3,4\}$, from which KL iterations yield a five-group $n_g = \{4,4,4,4,4\}$ partition ($Q \doteq 0.5886$, less than $0.5\%$ lower than the optimal). (\textit{For this example only}, we order the numbers in $n_g$ spatially along the line segment; we will typically use $n_g$ to indicate community sizes without spatial reference.)  Yet again, subnetwork-restricted partitioning provides an improvement, yielding an $n_g=\{2,2,3,3,3,3,2,2\}$ partition from which KL iterations converge either to the optimal $\{5,5,5,5\}$ partition or the nearly-optimal $\{4,4,4,4,4\}$ partition (depending on a random tie-breaking step). Another way out of this situation would be to generalize the implementation to allow forking and/or backtracking along the selected partitioning steps.  However, the potential downside is that this would allow so many choices that such an algorithm would become untenable on networks of any reasonable size.

\subsection{United States Congress Roll Call Votes}\label{sec:roll call}

To provide an interesting real-world example, we infer networks from Congressional roll call votes (obtained from Voteview \cite{voteview}) based on voting similarity between legislators, which is
determined according to the votes they cast (without using any
political information about the content of the bills on which they
voted). Noting that the definition of voting similarity is
definitively not unique, we choose to define the weighted link
between two legislators as the tally of the number of times they
voted in the same manner on a bill (i.e., either both for it or both against it)
divided by the total number of bills on which they both voted
(thereby accounting for abstentions and absences) during each
two-year Congressional term. For the purposes of studying voting similarities as a network between legislators, we ignore the perfect similarity of a legislator with herself, setting $A_{ii}=0$.

The purpose of this discussion is not an exhaustive political
analysis (which we present in Ref.~\cite{waugh}).  Instead, we aim only to show
examples in which the spectral partitioning methods we have proposed
are particularly advantageous. We therefore ignore for the time
being a number of important issues about the construction of the
voting-similarity networks and the selection of the null model. For
instance, given the dense nature of the inferred similarity network
and the relative uniformity in the total edge strength distributions
of the nodes, one might reasonably consider a uniform null
model \cite{spinglass} instead of Newman-Girvan modularity, and
it might also be insightful to explore resolution parameters
\cite{santo,spinglass} using either null model.  Moreover, the
entire similarity construction might be reasonably replaced
by signed networks that account separately for the
agreements and disagreements between legislators (using an
appropriate null model such as that discussed in Ref.~\cite{signed},
which is also compatible with the partitioning methods proposed in the present
manuscript).

Most of the highest modularity partitions identified in the (two-year) voting similarity networks of the House of Representatives and Senate across Congressional history consist of two communities
corresponding closely but not perfectly to the two major parties of
the day. This is similar to previous findings on legislative cosponsorship networks, for which all
of the Houses in the available data attain their highest modularity
partition after the initial bipartition, corresponding closely to a
Democrat/Republican split \cite{yan}. Although no direct tripartitioning steps
were used in Ref.~\cite{yan}, the subnetwork-restricted partitioning
(discussed in Section \ref{sec:implement}) was frequently able to
identify splits in the communities that appeared to include groups
of Southern Democrats and Northeastern Republicans who were not as
tightly tied to their respective parties. Such results are
consistent with political theories and observations about
low-dimensional legislative policy spaces \cite{pr,voteview},
including the assertion that the essentially one-dimensional
(Left--Right) legislative spectrum typically observed in U.~S.
history has not held as strongly during times when issues related to
slavery and civil rights have been of high legislative importance.

As an example of how this extra dimension in policy space can affect
the detected communities, we use the tools of the present manuscript
to explore the roll call voting similarity network of the 85th House
of Representatives (1957--1958), which passed the Civil Rights Act
of 1957. This network includes all 444 Representatives who voted
during the period (including midterm replacements). All but 386 of
the 98346 pairs of legislators have non-zero similarity weight in
this network, with 251 of the empty similarities affiliated with
House Speaker Sam Rayburn [D-TX], who chose to vote ``present''
(treated as an abstention) in all but one vote, in which he broke a tie
on an amendment to the Interstate Commerce Act.

Recursive bipartitioning of the 85th House identifies a partition
with two communities of sizes $n_g=\{230,214\}$ (with $Q \doteq 0.07935$) and
subsequent KL iterations do not result in a higher-modularity partition.
As expected, such communities correspond reasonably but not
perfectly with party affiliation; the Republican community includes
16 Democrats, and the Democratic community includes 6 Republicans.  Direct tripartioning yields three communities of sizes $n_g=\{192,187,65\}$ and a higher modularity ($Q \doteq 0.08019$).
Subsequent KL iterations yield a $\{197,190,57\}$ partition with
$Q \doteq 0.08063$, which is the highest modularity we identified in this
network using the methods of the present manuscript \footnote{This is the same best partition identified by restricting attention to the so-called ``non-unanimous'' votes \cite{waugh}, though of course the corresponding value of modularity obtained is then slightly different because the
underlying edge-weight similarities are different.}.  Whether or not
one allows tripartitions during a single step, the same partition is
obtained by subnetwork-restricted partitioning followed by KL
iterations (though obviously by a more convoluted process in this
case). The largest of these three communities includes 196
Republicans (all but 8 of them) and Democratic Speaker Sam Rayburn, the ``misplacement'' of whom arises from his having cast only the one aforementioned vote. The next largest community is dominated by Democrats and includes the remaining 8 Republicans. The smallest community consists entirely of Democrats from Southern states.

To demonstrate the use of these methods with a quality function other
than modularity, we also studied voting similarities using
the quality function with additional self loops proposed by Arenas \emph{et
al}.~\cite{arenas08}. As emphasized earlier, any quality function
that can be expressed with a $\mathbf{B}$ matrix as in
\eqref{modularity} is amenable to these spectral partitioning methods. Adding such self-loops of weight $r$ to the voting
similarity adjacency matrix and null model \footnote{The appearance
of self-loops in the adjacency matrix $\mathbf{A}$ changes the
optimum partition through the corresponding changes in node
strengths $k_i$ and total edge weight $m$ in the null model
$\mathbf{P}$.}, our procedure identifies the same three-community
structure in the range $-3.4 \lesssim r \lesssim 4.9$ (with $r = 0$
corresponding to the usual definition of modularity). Below this range,
the algorithm identifies a two-community partition; above this range, it identifies a four-community partition. We note that the use of the quality function with self-loops demonstrates a distinct downside in using subnetwork-restricted partitioning, as even modest positive values
of $r$ lead such extended subpartitioning to proceed all the way
down to single-node groups, thereby utilizing significantly greater
computing time without leading to any improvement in final
communities than would be obtained by KL iterations
starting simply from individual-node groups. We therefore do not
advocate subnetwork-restricted partitioning for any resolutions
significantly finer than that corresponding to the traditional
definition of modularity.

\subsection{Network Coauthorship}\label{coauth}

The graph of coauthorships in network science publications \cite{newman2006pre}
has become a well-known benchmark example in the network science
community. The largest connected component of this network consists
of 379 nodes, representing authors, with 914 weighted edges
indicating the coauthored papers between pairs of scientists.

Without the use of tripartitioning steps, spectral recursive
bipartitioning yields a $Q \doteq 0.8188$ partition with 29 communities of various sizes. KL
iterations starting from this state yields a $Q \doteq 0.8409$ partition
with 27 communities (subject to random tie-breaking instances). In this example, the subnetwork-restricted
modularity maximization does \textit{not} improve the collection of
communities that one obtains; rather, it generates a partition of
120 subcommunities that KL iterations subsequently merge into a
$Q \doteq 0.8190$ partition with 37 communities. That is, while the
modularity obtained in this manner is better than that from the
original recursive bipartitioning, it is not as good as
that obtained when KL iterations are used directly after the
modularity-increasing bipartitioning terminates.  In contrast, using KL
iterations that allow new groups to be created proceeds without ever generating such new groups when started from the 29-group partition obtained by recursive
bipartitioning (even with formation of new groups selected in any tie breaking).  Hence, we note that subnetwork-restricted modularity
maximization yields better results (i.e., a higher-modularity partition) in some cases, whereas KL iterations that allow the creation of new groups yield better results in others.

When tripartitioning steps are used along with bipartitioning, one obtains a final
partition (after KL iterations) with a higher modularity, even
though some of the intermediate states have lower modularity. Taking
the best available division at each stage, the algorithm first
splits the network into three groups.  Each of these is then further
divided three ways, eventually yielding a $Q \doteq 0.8032$ partition
with 39 communities. Even though this modularity is lower than that
obtained using spectral bipartitioning by itself, applying KL
iterations at this stage yields a $Q \doteq 0.8427$ partition with 24
communities, which is slightly better than the best result described above.
In this case, too, the subnetwork-restricted modularity maximization
plus KL iterations yields a final partition with lower modularity
($Q \doteq 0.8220$).  Such results highlight an important point:
Using multiple combinations of these methods might give better results
than fixating on any specific combination.

\begin{figure}
    \centering
    \includegraphics[width=.45\textwidth]{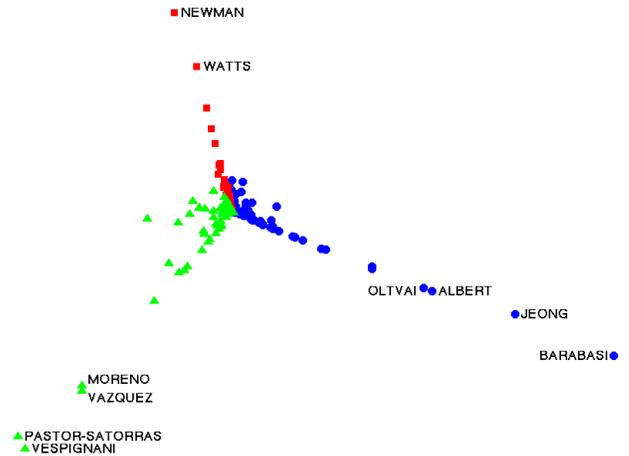}
    \caption {(Color online) Two-dimensional vertex vector coordinates of the 379-node largest connected component of the network scientists coauthorship graph.  Colors/shapes denote the post-KL three-community partition discussed in the text.}
\label{fig : coaut 1}
\end{figure}

The initial tripartitioning of the full network is itself
interesting as an example of our divide-and-conquer approach,
as it illustrates the extent to which KL iterations can improve a
spectrally-obtained partition. It also provides a compelling
visualization (see Fig.~\ref{fig : coaut 1}) of the three-way
division of the vertex vectors in the plane. The partition obtained
after the initial split contains groups of sizes
$\{164, 118, 97\}$ with $Q \doteq 0.5928$. Applying KL iterations to this
partition moves 54 of the nodes and gives a $Q \doteq 0.6354$ partition (see Fig.~\ref{fig : coaut 1}) with groups of sizes $\{136, 128, 115\}$.  (Note that, in contrast to this example, we do not typically apply KL iterations after each recursive subdivision step;
instead, we apply them after exhausting spectral techniques.)  Although
KL iterations moved a significant fraction of the nodes, the regions
in Fig.~\ref{fig : coaut 1} nevertheless appear to resemble
non-overlapping wedges because the nodes that were moved are
all too close to the origin to visualize the overlap induced by the
KL iterative improvement. For convenience, we have indicated in Fig.~\ref{fig
: coaut 1} the last names of the 10 authors (some of whom are rather familiar) with largest vertex
vector magnitudes.  Finally, although this three-community
partition illustrates some of the well-known research camps in
network science, it is important to remember that the
modularity-maximizing partition of this network has many more than
three communities.

\section{Conclusions} \label{sec:conclusions}

We have presented a computationally-efficient method for spectral tripartitioning of a network using the leading pair of eigenvectors of a modularity matrix.  
Our algorithm, which can be applied without modification to a broad class of quality functions, extends the previously-available methods for spectral
optimization of modularity in Refs.~\cite{newman2006pnas,newman2006pre}. Paired with
spectral bipartitioning in a recursive implementation with
subsequent KL iterations and an optional subnetwork-restricted modularity
maximization extension, the inclusion of possible tripartitioning
steps significantly expands the possible partitions that can be
efficiently considered in the heuristic optimization of modularity or any of the wide class of quality functions that can be expressed in similar matrix form (such as those that generalize modularity
using a multiplicative resolution parameter
\cite{spinglass} or self-loops \cite{arenas08}).

Our investigation also provides an important cautionary tale about
community detection in networks.  Despite a wealth of recent
research on this subject, it sometimes remains unclear how to
interpret the results of graph partitioning methods and which
methods are most appropriate for which particular data sets
\cite{santo,facebook,santobig,ourreview}. While recursive subdivision seems to give
some hierarchical information about network structure and how nodes
are grouped (see Section~\ref{sec:examples} and
Refs.~\cite{structpnas,bcs,congshort,conglong,conggallery}), the
hierarchies that one obtains might indicate as much about the
algorithms employed as they do about any  true hierarchical
structures of communities (see the discussion in
Ref.~\cite{clausetnature}). Moreover, it is important to stress that
the process of always taking the best modularity at each divisive
step can lead to states that are not as good as might have been
obtained from other choices in the forking decision process, and
that the best post-KL partitions are not always obtained from the
best available pre-KL states. Because of the necessarily
heuristic nature of trying to obtain a high quality partition in
polynomial time, it is beneficial to have access to a variety of
computationally-efficient tools with which to explore the complicated landscape of possible community partitions.

\begin{acknowledgments}

We thank Diana Chen, Ye Pei, Stephen Reid, Amanda Traud, and Yan
Zhang for helping to develop some of the code used for exploring
some of the ideas in this paper.  We also thank Keith Poole, Andrew
Waugh, and an anonymous referee for useful comments.  We obtained
the Congressional roll call voting data from Keith Poole's Voteview
\cite{voteview} and the network science coauthorship data from Mark
Newman.  TR's contribution to this project was supported in part by
a Summer Undergraduate Research Fellowship (SURF) award at the
University of North Carolina at Chapel Hill (UNC).  Additional support for PJM and TR was obtained from the National Science Foundation (DMS-0645369) and from start-up funds provided to PJM by the Department of Mathematics and the Institute for Advanced Materials, Nanoscience and Technology at UNC.

\end{acknowledgments}



\end{document}